\documentclass{INTERSPEECH2023}

\interspeechcameraready 

\usepackage{amsmath}

\title{Diff-E: Diffusion-based Learning for Decoding Imagined Speech EEG}
\name{Soowon Kim$^1$, Young-Eun Lee$^2$, Seo-Hyun Lee$^2$, Seong-Whan Lee$^1$}
\address{$^1$Department of Artificial Intelligence,  Korea University, Republic of Korea \\
  $^2$Department of Brain and Cognitive Engineering, Korea University, Republic of Korea}
\email{soowon\_kim@korea.ac.kr, ye\_lee@korea.ac.kr, seohyunlee@korea.ac.kr, sw.lee@korea.ac.kr}

\begin{document}

\maketitle

\begin{abstract}
Decoding EEG signals for imagined speech is a challenging task due to the high-dimensional nature of the data and low signal-to-noise ratio. In recent years, denoising diffusion probabilistic models (DDPMs) have emerged as promising approaches for representation learning in various domains. Our study proposes a novel method for decoding EEG signals for imagined speech using DDPMs and a conditional autoencoder named Diff-E. Results indicate that Diff-E significantly improves the accuracy of decoding EEG signals for imagined speech compared to traditional machine learning techniques and baseline models. Our findings suggest that DDPMs can be an effective tool for EEG signal decoding, with potential implications for the development of brain-computer interfaces that enable communication through imagined speech.
\end{abstract}

\section{Introduction}
Speech is a crucial mode of communication for humans, allowing us to convey complex ideas and thoughts to others through sound patterns \cite{lieberman1991uniquely, peters2012speaking, lieberman2007evolution}. It is a fundamental part of our social and cultural interactions, enabling us to build relationships, share information, and express emotions. Speech is a highly complex process that involves the coordinated activity of various brain regions and muscles, including the larynx, tongue, lips, and jaw. Unfortunately, several individuals cannot speak due to physical limitations such as locked-in syndrome, a condition where they are fully conscious but unable to move or communicate verbally \cite{laureys2005locked, houde2011speech}. 
The inability to speak can significantly impact an individual's quality of life, as it limits their ability to interact with others and express themselves effectively. Therefore, developing effective methods for restoring communication in individuals with locked-in syndrome and other speech impairments is a critical area of research. In this context, our research explores decoding brain signals as a potential method for enabling silent communication \cite{chaudhary2016brain, birbaumer2008brain, chaudhary2017brain}. 


Electroencephalography (EEG) is a non-invasive technique for measuring the brain's electrical activity. EEG signals are acquired by placing electrodes on the scalp, which detect the electrical activity generated by the brain's neurons. EEG signals have been widely used in neuroscience research to study brain function, communication, and medical applications such as diagnosing neurological disorders. In recent years, EEG decoding has gained increasing attention to translate EEG signals into meaningful information, such as speech or attention detection \cite{cai2021eeg, su2022stanet, lee2023towards, lee2022eeg}. 
EEG decoding for speech processing is a particularly challenging task, as it involves decoding complex and highly variable neural patterns associated with speech production and perception \cite{li2021biologically}. 
Moreover, EEG signals are often contaminated with noise and artifacts, which can affect the accuracy of the decoding process \cite{lee2020real}. Therefore, developing accurate and robust methods for decoding EEG signals associated with imagined speech is an active area of research, with potential applications in speech rehabilitation, human-computer interaction, and other related fields.

\begin{figure}[t]
  \centering
  \includegraphics[width=\linewidth,height=\textheight,keepaspectratio]{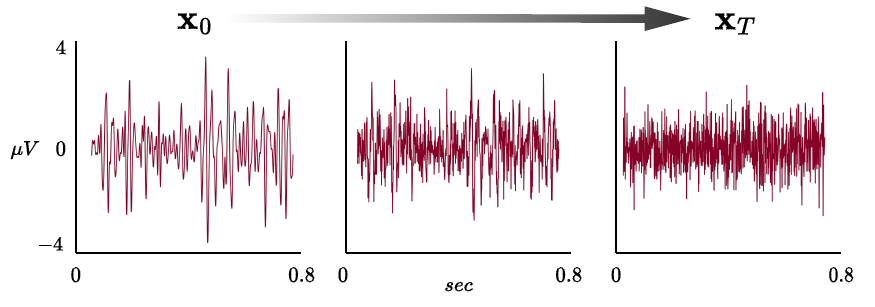}
  \caption{ The figure illustrates the forward diffusion process applied to a section of the recorded EEG signal captured from Broca's area, FT7, of Subject 1.}
  \label{fig:diffusion_forward}
\end{figure}

Denoising diffusion probabilistic models (DDPMs) have become a powerful tool for identifying sophisticated patterns within complex data, especially when dealing with high-dimensional data \cite{ho2020denoising, song2020score, dhariwal2021diffusion}. For DDPMs, the forward diffusion process introduces noise to the original data, transforming it into a corrupted version. This process is modeled by adding Gaussian noise at each time step, effectively simulating a random walk in the latent space. The aim of the denoising diffusion model is then to reverse this process, i.e., to denoise the corrupted data and recover the original signal.
In recent years, DDPMs have gained attention as a promising approach for time-series data processing and analysis as well, including audio and video \cite{tashiro2021csdi, rasul2021autoregressive}. In the context of EEG decoding, DDPMs may offer a promising approach for capturing the complex relationships between neural activity and language processing during imagined speech tasks. This can enable accurate decoding of EEG signals associated with imagined speech, which has potential applications in various fields such as speech rehabilitation and neuroscience.

DDPMs have also been employed in representation learning. Preechakul et al. \cite{preechakul2022diffusion} demonstrated that diffusion autoencoders can yield a more interpretable and decodable latent data representation than conventional autoencoders. Zhang et al. \cite{zhang2022unsupervised} displayed the effectiveness of pre-trained diffusion models using high-dimensional data representations. Their approach involves a trainable model that predicts a mean shift based on the encoded representation. It is trained to bridge as much of the gap as possible, a gap that emerges during the forward process of DDPMs. Consequently, the encoder is compelled to extract as much information as it can from images to assist with this bridging process.
In this study, building upon prior research, we present a novel approach to decode EEG signals using DDPMs and a Conditional Autoencoder (CAE). The CAE aids in learning meaningful features potentially lost during the forward process in DDPMs. Furthermore, we've integrated a jointly-trained classifier to improve decoding performance. To the best of our knowledge, this proposed approach represents the first attempt to utilize diffusion models to decode EEG signals associated with imagined speech.


\section{Related Works}
Decoding EEG signals using deep learning approaches is a challenging problem due to various factors, including the scarcity of data, poor signal-to-noise ratio, and high inter- and intra-individual variability \cite{tayeb2019validating}. Despite these challenges, several studies have explored different EEG decoding techniques for various applications, including speech decoding.


Schirrmeister et al. \cite{schirrmeister2017deep} developed DeepConvNets for end-to-end learning of EEG signals in human subjects. They utilized machine learning techniques such as batch normalization and exponential linear units and a cropped training strategy to improve decoding performance, achieving accuracy levels similar to filter bank common spatial patterns algorithms. Another study by Lawhern et al. \cite{lawhern2018eegnet} proposed EEGNet, a compact CNN architecture specifically designed for EEG classification tasks. EEGNet uses depthwise and separable convolutions, which can extract features tailored to the unique characteristics of EEG control signals. The model has been tested on four BCI paradigms and shown to accurately classify EEG signals while being compact in terms of the number of parameters in the model.

Lee et al. \cite{lee2020neural} investigated the underlying characteristics affecting the decoding performance of two emerging BCI paradigms: imagined speech and visual imagery. The authors analyzed EEG signals filtered in six frequency ranges for 13 class classifications, including a rest class. They identified cortical regions relevant to both paradigms using semantic levels of word properties according to their concepts. The study showed high accuracy rates across all classes and multi-class scalability in both paradigms, providing crucial information for improving the efficiency of BCI applications.

Diffusion-based methods for time-series data have been extensively researched recently. One such method proposed in \cite{alcaraz2022diffusion} is based on a structured state space model that incorporates a diffusion process to capture the underlying dynamics of the data. This method is able to impute and forecast missing values in time series data by utilizing the estimated diffusion process. The authors demonstrate the effectiveness of their approach on various real-world datasets and show that it outperforms existing methods for time-series imputation and forecasting. Jeong et al. \cite{jeong2021diff} propose a method for improving the quality of synthetic speech generated by TTS systems. The proposed method utilizes DDPMS to remove noise from the intermediate representations of speech generated by TTS systems. By reducing noise in the intermediate representations, the proposed method is able to produce clearer and more natural-sounding speech. The authors evaluate the effectiveness of their approach on several real-world datasets and demonstrate that it outperforms existing methods for improving the quality of synthetic speech.


\section{Materials and Methods}
\begin{figure}[t]
  \centering
  \includegraphics[width=\linewidth,height=\textheight,keepaspectratio]{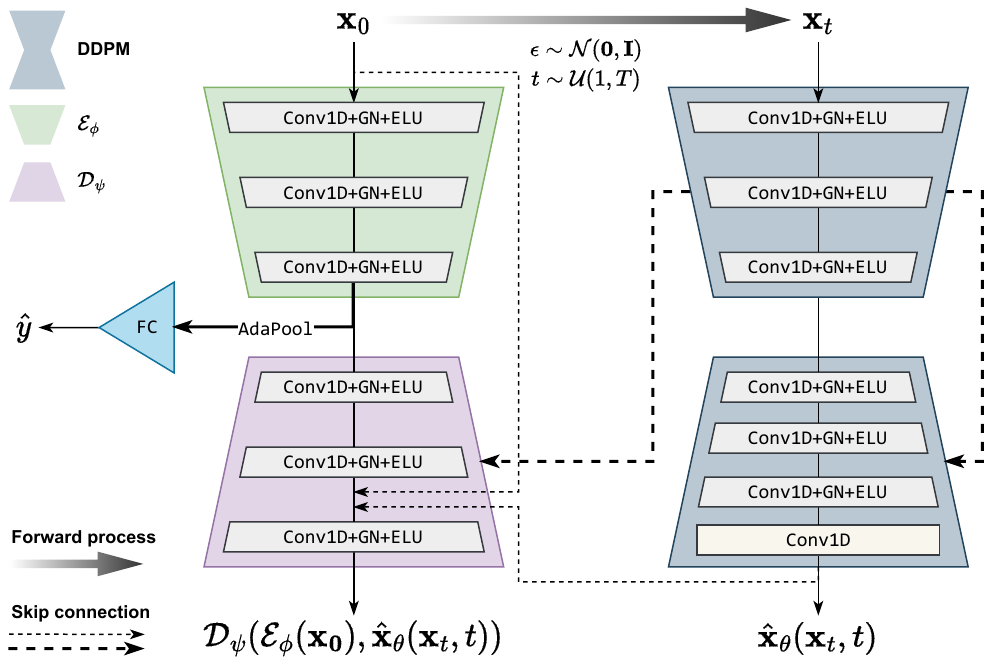}
  \caption{A flowchart of Diff-E for EEG signal decoding. The nodes labeled as \texttt{Conv1D}, \texttt{GN}, and \texttt{Linear} represent one-dimensional convolutional layer, group normalization, and linear layer, respectively. The node labeled as \texttt{ELU} represents a nonlinear activation function. Adaptive average pooling layer, \texttt{AdaPool} is used to make input for the classifier, \texttt{FC}.}
  \label{fig:method}
\end{figure}


\subsection{Denoising Diffusion Models}
DDPMs are a type of machine learning model that can learn complex probability distributions over data. The "forward process" in DDPMs is determined by a fixed Markov chain that progressively introduces Gaussian noise to the data. The forward process in DDPMs starts with a probability distribution called $q(\mathbf{x}_0)$, which represents the uncorrupted original data. This distribution is then transformed iteratively using a sequence of Markov diffusion kernels called $q(\mathbf{x}_t|\mathbf{x}_{t-1})$, which are Gaussian with a fixed variance schedule $\{\beta_t\}^T_{t=1}$. The process can be expressed by:

\begin{equation}
    q(\mathbf{x}_t|\mathbf{x}_{t-1})=\mathcal{N}(\mathbf{x}_t;\sqrt{1-\beta_t}\mathbf{x}_{t-1},\beta_t\mathbf{I})
\end{equation}
\hspace{0.5cm}
\begin{equation}
    q(\mathbf{x}_{1:T}|\mathbf{x}_0)=\prod_{t=1}^{T}q(\mathbf{x}_t|\mathbf{x}_{t-1})
\end{equation}

Diffusion process with Gaussian noise can corrupt the original data with arbitrary stage, $t$ with the notation $\alpha_t=1-\beta_t$ and $\Bar{\alpha}_t=\prod_{s=1}^{t}\alpha_s$:

\begin{equation}
    q(\mathbf{x}_t|\mathbf{x}_0)=\mathcal{N}(\mathbf{x}_t;\sqrt{\Bar{\alpha}_t}\mathbf{x}_0, (1-\Bar{\alpha}_t) \mathbf{I})
\end{equation}

Ho et al. \cite{ho2020denoising} proposed a method to train a model that takes a noisy sample $\mathbf{x}t$ and predicts its noise by training a network $\epsilon\theta(\mathbf{x}_t,t)$. In contrast, our study trains Diff-E to predict the uncorrupted original signal, $\mathbf{x}_0$, instead of predicting the injected noise:

\begin{equation}
\mathcal{L}_{\text{DDPM}}(\theta) = ||\mathbf{x}_0 - \hat{\mathbf{x}}_\theta(\mathbf{x}_t,t)||
\label{eq:DDPM}
\end{equation}

Here, $\theta$ are the DDPM parameters, and $\hat{\mathbf{x}}_\theta(\mathbf{x}_t,t)$ is the prediction of the DDPM. We randomly sample timestep $t$ from a uniform distribution, $\mathcal{U}(1, T)$. The objective function encourages the model to denoise the noisy input and produce output that is close to the original input signal. We have adopted a time-conditional UNet architecture \cite{ronneberger2015u} as described in Fig. \ref{fig:method}, similar to the model used in the original paper \cite{ho2020denoising}, with modifications to adapt it for EEG data.

\subsection{Conditional Autoencoder}
Similar to the work \cite{zhang2022unsupervised}, the CAE aims to derive meaningful representations of EEG signals by leveraging the errors that occur during the reconstruction process by the DDPM. The forward pass of the DDPM causes information loss, which the CAE attempts to compensate for by identifying and rectifying these errors. This enables the CAE to generate more accurate representations of the original EEG signals. Specifically, we employ the objective function for the CAE as follows:
\begin{equation}
    \mathcal{L}_{\text{CAE}}(\psi, \phi) = ||\mathcal{L}_{\text{DDPM}}(\theta) -\mathcal{D}_\psi(\mathcal{E}_\phi(\mathbf{x}_0),\hat{\mathbf{x}}_\theta(\mathbf{x}_t, t))||,
\label{eq:CAE}
\end{equation}

The CAE is composed of an encoder network, denoted by $\mathcal{E}_\phi$, and a decoder network, denoted by $\mathcal{D}_\psi$, which is similar in structure to the DDPM. However, $\mathcal{D}_\psi$ receives skip connections not from the outputs of $\mathcal{E}_\phi$, but from the DDPM layers. This enables $\mathcal{D}_\psi$ to be implicitly conditioned on the corruption stage of the forward process of the DDPM, as shown in Fig. \ref{fig:method} as \textit{dashed arrows}. Furthermore, to improve the reconstruction of $\mathcal{L}_{\text{DDPM}}$, the original signal, $\mathbf{x}_0$, and the output of the DDPM, $\hat{\mathbf{x}}_\theta(\mathbf{x}_t,t)$, are also skip-connected as inputs to the last layer of $\mathcal{D}_\psi$ as depicted in Fig. \ref{fig:method} as \textit{thin dashed arrows}.

\subsection{Classifier}
After $\mathcal{E}_\phi$ processes the data, the output is transformed into a compact representation, $\mathbf{z}$, by collapsing the time dimension into a single dimension. This is achieved by applying an adaptive average pooling layer, which effectively creates a one-dimensional latent vector. The compressed representation is then fed into the linear classifier, $\mathcal{C}_\rho$, which is jointly trained with the CAE to further separate the representations of each class and perform the classification task. The dimension of $\mathbf{z}$ is constant at $256$ throughout the experiment. To include the classification loss in the objective function of the CAE, we modified it as follows to become the overall Diff-E objective:

\begin{equation}
    \begin{split}
        \mathcal{L}_{\text{Diff-E}}(\psi, \phi, \rho) &= ||\mathcal{L}_{\text{DDPM}}(\theta) -\mathcal{D}_\psi(\mathcal{E}_\phi(\mathbf{x}_0),\hat{\mathbf{x}}_\theta(\mathbf{x}_t, t))||\\
        &+\alpha||\hat{y}-y||_2
    \end{split}
\label{eq:Diff-E}
\end{equation}

where $\hat{y}=\mathcal{C}_\rho(\mathbf{z})$. Here, $\rho$ is adjustable parameters for $\mathcal{C}_\rho$, and $y$ is the true label of the input signal. The hyperparameter $\alpha$ governs the relative importance of the reconstruction loss and the classification loss. $0.1$ is chosen for the value of $\alpha$ throughout the experiment. During the inference, only $\mathcal{E}_\phi$ and $\mathcal{C}_\rho$ are used to classify the signals. Specifically, the predicted label is obtained as  $\hat{y}=\mathcal{E}_\phi(\mathcal{C}_\rho(\mathbf{x}_0))$.

In order to evaluate the effectiveness of Diff-E, it is compared to other existing methods that have demonstrated their applicability to decoding EEG signals in various paradigms, such as motor imagery and event-related potentials. This comparison is conducted to assess the performance of Diff-E and to determine its suitability for imagined speech EEG signal decoding applications.

\subsection{Model Implementation Details}
The DDPM and CAE architectures in our study were designed with each layer comprising a one-dimensional convolution (\texttt{Conv1D}), group normalization (\texttt{GN}), and activation function (\texttt{ELU}). In $\mathcal{E}_\phi$, adaptive average pooling (\texttt{AdaPool}) is applied to transform the output into a compressed representation, $\mathbf{z}$. The total number of trainable parameters for both DDPM and CAE was approximately 300,000, while the linear classifier had 400,000 trainable parameters.

We trained the models using the RMSProp optimizer with a fixed random seed of 42 and a cyclic learning rate scheduler developed by Smith \cite{smith2017cyclical}. The base learning rate was set to $9*10^{-5}$, and the maximum learning rate was set to $1.5*10^{-3}$. The training process was performed over a total of 500 epochs. During training, we used L1 loss for both the DDPM and CAE components of the model. We also used the mean squared error loss against one-hot encoded labels for the linear classifier classification task. To evaluate the performance of the model, we set aside 20\% of the data for testing with a fixed random seed. Our implementation is available at \url{https://github.com/yorgoon/DiffE/}.

\subsection{Dataset}
\subsubsection{Data Description}
The data utilized in this study was obtained from a prior study conducted by Lee et al. \cite{lee2020neural}. The study involved recruiting twenty-two healthy participants, of whom fifteen were male, with a mean age of 24.68 ± 2.15. None of the participants had a history of neurological disease or language disorder, and they did not have any hearing or visual impairments. Additionally, they refrained from taking drugs for twelve hours before the session. All participants had received high-quality English education for more than 15 years. In the imagined speech task, twenty-two subjects were instructed to imagine saying twelve different words or sentences, such as ``ambulance," ``clock," ``hello," ``help me," ``light," ``pain," ``stop," ``thank you," ``toilet," ``TV," ``water," and ``yes," as well as a resting state, resulting in a total of thirteen classes that a communication board commonly used in hospitals for patients with paralysis or aphasia. These words are considered to be essential for patient communication. The researchers used a 64-channel EEG cap with active Ag/AgCl electrodes that followed the international 10-10 system to record the EEG signals. The FCz and FPz channels were set as the reference and ground electrodes, respectively. Brain Vision/Recorder software (BrainProduct GmbH, Germany) was used to collect the EEG signals, which were then operated using MATLAB 2018a software. The researchers ensured that the impedance of all electrodes was maintained below $10k\Omega$. The researchers randomly presented twenty-two blocks of twelve words and a rest class. Each of the twenty-two participants contributed one thousand three hundred samples, consisting of one hundred samples per category. The study was approved by the Korea University Institutional Review Board [KUIRB-2019-0143-01] and was conducted in accordance with the Declaration of Helsinki.

\subsubsection{Preprocessing}
In this study, we employed several preprocessing steps to ensure the quality and accuracy of the EEG data. The first step was to apply band-pass filtering to the signals between 0.5 and 125Hz, with additional notch filtering at 60 and 120Hz to eliminate power line interference. The data was then re-referenced using a common average reference method to reduce the impact of any noise present in the data. To eliminate ocular and muscular artifacts that can be caused by movement or sounds, we used automatic electrooculography and electromyography removal methods. After removing the artifacts, the EEG signals were selected in the high-gamma frequency band for training the model and data analysis. The dataset was then epoched into 2-second segments, with baseline correction applied 500ms before the task. All of the preprocessing steps were carried out using MATLAB-based tools, including the OpenBMI Toolbox \cite{leeMH2019eeg,lee2021decoding}, BBCI Toolbox \cite{krepki2007berlin}, and EEGLAB \cite{Delorme2004EEGLAB}.

\section{Results and Discussion}
\subsection{Performance Comparison}

In this study, we compared the performance of our proposed Diff-E method with three other methods, DeepConvNet \cite{schirrmeister2017deep}, EEGNet \cite{lawhern2018eegnet}, and Lee et al. \cite{lee2020neural}, for EEG signal decoding. The results are presented in Table \ref{tab:result}. Diff-E outperformed all other methods, including DeepConvNet, EEGNet, and Lee et al., in both accuracy and area under the curve (AUC). The average accuracy of Diff-E was 60.63\%, with a standard deviation of 8.03\%, and the average AUC was 90.39\%, with a standard deviation of 3.48\%. Notably, Diff-E achieved significantly higher accuracy and AUC than the other methods. The average accuracy of DeepConvNet, EEGNet, and Lee et al. were 21.00\%, 32.78\%, and 46.27\%, respectively. The average AUC of DeepConvNet, EEGNet, and Lee et al. were 69.55\%, 78.07\%, and 79.97\%, respectively. These findings highlight the superior performance of our proposed Diff-E method for EEG signal decoding.

Our study's results were unexpected since commonly used, and well-known methods such as EEGNet and DeepConvNet for EEG decoding underperformed compared to a traditional machine-learning method, such as a common spatial pattern-support vector machine method used by Lee et al. \cite{lee2020neural}. These methods are frequently used for decoding EEG signals from various paradigms, including motor imagery and event-related potentials. Our findings emphasize the importance of carefully selecting the appropriate model architecture based on the task at hand, the EEG paradigms used, and other relevant factors.



\begin{table}[t]
\centering
\renewcommand{\tabcolsep}{3mm}
\caption{Accuracy and AUC scores for imagined speech classification}
\label{tab:result}
{
\begin{tabular}{lcc}
\hline
\textbf{Subject}     & \textbf{Accuracy (\%)} & \textbf{AUC (\%)}\\ 
\hline
\textbf{DeepConvNet} & 21.00 $\pm$ 4.27 & 69.55 $\pm$ 3.51 \\
\textbf{EEGNet}      & 32.78 $\pm$ 4.98 & 78.07 $\pm$ 4.27 \\
\textbf{Lee et al.\cite{lee2020neural}} & 46.27 $\pm$ 6.79 & 79.97 $\pm$ 4.83 \\
\textbf{Diff-E} & \textbf{60.63 $\pm$ 8.03} & \textbf{90.39 $\pm$ 3.48} \\
\hline
\end{tabular}
}
\end{table}

\begin{table}[t]
\centering
\caption{Ablation study}

{
\begin{tabular}{lcc}
\hline
\textbf{Components}     & \textbf{Accuracy (\%)}  & \textbf{AUC (\%)}\\ 
\hline
\textbf{Diff-E} & \textbf{60.63  $\pm$ 8.03} & \textbf{90.39 $\pm$ 3.48} \\
\textbf{w/o DDPM} & 46.59 $\pm$ 8.35 & 86.54 $\pm$ 5.36 \\
\textbf{w/o DDPM \& $\mathcal{D}_\psi$} & 47.01  $\pm$ 7.80 & 46.27 $\pm$ 6.79 \\

\hline
\end{tabular}%
}
\label{tab:ablation}
\end{table}

\subsection{Ablation Study}
To evaluate the impact of each component of our proposed method, we performed an ablation study. The study involved removing the diffusion process and training the CAE's encoder and classifier separately to assess the effects of learning from the DDPM's errors and whether the autoencoder-style approach is advantageous for learning meaningful representations. We aimed to determine the significance of each component of Diff-E. During the training process, the CAE was trained only to reconstruct the original signals, and a linear classifier was used without the use of DDPM. Additionally, we conducted another ablation study solely on the $(\mathcal{E}_\phi,\mathcal{C}_\rho)$ component for the classification task, which is essentially a CNN-based classifier. 

As shown in Table \ref{tab:ablation}, the results indicate that Diff-E outperformed the other models, and removing the DDPM or solely using $(\mathcal{E}_\phi,\mathcal{C}_\rho)$ led to a significant decrease in accuracy. These results indicate that without DDPM, CAE alone does not offer any advantage over traditional machine learning techniques.

\subsection{Future Works}

To enhance the effectiveness and robustness of our proposed method for EEG signal decoding, we intend to address several limitations in our future work. Our method was tested on a non-public dataset, as it had larger and more diverse classes than other public datasets, which usually have only a few classes \cite{coretto2017open}. Although our EEG dataset was relatively large, it may still be inadequate for large deep-learning models to optimize effectively. To improve the applicability of our method, we plan to collect larger datasets with more subjects and apply our method to various EEG paradigms, including motor imagery and event-related potentials. This will offer a more comprehensive evaluation of our method's effectiveness and improve its robustness across different paradigms. Additionally, we plan to investigate methods for end-to-end learning directly from raw EEG signals, potentially reducing the need for extensive pre-processing and enhancing the efficiency of our method. 

\section{Conclusion}

This work contributes to the growing body of research on deep-learning-based approaches to EEG signal decoding, specifically for the task of imagined speech. The use of DDPMs, a relatively new approach in this field, provides an innovative solution that has demonstrated superior performance compared to traditional methods such as DeepConvNet and EEGNet. The results of our study suggest that DDPMs have the potential to become a valuable tool in EEG signal processing and can be further explored for other applications in this domain. Our study also highlights the importance of selecting appropriate model architectures for different EEG paradigms. This consideration is crucial to achieving high accuracy and performance in EEG signal decoding, as demonstrated by the inferior performance of DeepConvNet and EEGNet in our study. The selection of an appropriate model architecture should take into account various factors, such as the characteristics of the EEG data and the task at hand. Overall, our study provides a promising approach to accurately decoding EEG signals related to imagined speech, which can have significant implications in the development of brain-computer interfaces for communication and assistive technology. This work also contributes to advancing the understanding and utilization of deep-learning-based approaches in EEG signal processing and decoding.



\section{Acknowledgements}
\thanks{This work was supported by Institute for Information \& Communications Technology Planning \& Evaluation (IITP) grant funded by the Korea government (MSIT) (No.2021-0-02068, Artificial Intelligence Innovation Hub; No. 2019-0-00079, Artificial Intelligence Graduate School Program (Korea University); No. 2017-0-00451, Development of BCI based Brain and Cognitive Computing Technology for Recognizing User’s Intentions using Deep Learning).}

\bibliographystyle{IEEEtran}
\bibliography{mybib}

\end{document}